\documentclass[prl,twocolumn,floatfix]{revtex4}
\newcommand{\EE}{{\cal E}^{\scriptscriptstyle \sf q}} 
\newcommand{\EEd}{\dot{\cal E}^{\scriptscriptstyle \sf q}} 
\newcommand{\EEE}{{\cal E}^{\scriptscriptstyle \sf o}} 
\newcommand{\BB}{{\cal B}^{\scriptscriptstyle \sf q}} 
\newcommand{\BBd}{\dot{\cal B}^{\scriptscriptstyle \sf q}}  
\newcommand{\BBB}{{\cal B}^{\scriptscriptstyle \sf o}} 
\begin{document}
\title{Metric of a tidally distorted, nonrotating black hole}  
\author{Eric Poisson}
\affiliation{Department of Physics, University of Guelph, Guelph,
Ontario, Canada N1G 2W1}
\date{March 8, 2005} 
\begin{abstract}
The metric of a tidally distorted, nonrotating black hole is presented
in a light-cone coordinate system that penetrates the event horizon and
possesses a clear geometrical meaning. The metric is expressed as an
expansion in powers of $r/{\cal R} \ll 1$, where $r$ is a measure of
distance from the black hole and ${\cal R}$ is the local radius of
curvature of the external spacetime; this is assumed to be much
larger than $M$, the mass of the black hole. The metric is calculated
up to a remainder of order $(r/{\cal R})^4$, and it depends on a
family of tidal gravitational fields which characterize the hole's
local environment. The coordinate system allows an easy identification
of the event horizon, and expressions are derived for its surface
gravity and the rates at which the tidal interaction transfers mass
and angular momentum to the black hole. 
\end{abstract}
\maketitle

{\it Introduction.} The numerical integration of the Einstein field
equations for the simulation of a two-body system of black holes is
currently the subject of intense research activity. This work
is largely motivated by ongoing attempts to measure gravitational
waves, an effort which requires the theoretical input of detailed
predictions for the expected waves. The numerical simulations
must proceed from initial data that correctly describe a situation of
astrophysical interest, a spacelike hypersurface that contains two
black holes in bound orbital motion, together with the gravitational
radiation that was emitted by the system in the recent past. To find
realistic initial data has proved a challenge \cite{cook}, and it has
been suggested \cite{PN,alvi} that these could be reliably constructed
using post-Newtonian (and post-Minkowskian) methods.  

While a post-Newtonian metric is adequate to describe the mutual  
gravity of a weakly-bound two-body system, it fails to describe the 
strong gravity of each black hole. Building on techniques developed by
previous authors \cite{thorne-hartle,death}, Alvi remedied this
situation \cite{alvi} by matching the post-Newtonian metric to two
perturbed versions of the Schwarzschild metric (each black hole being
tidally distorted by the other hole), in a buffer region in which both 
approximations are valid. Alvi's metric is ready to use in numerical
simulations, and in this Letter we improve on a part of Alvi's
construction by providing a more accurate version of the perturbed
Schwarzschild metric.    

We calculate the metric of a tidally distorted, nonrotating black hole 
to a level of accuracy that has never been attempted before. The hole
is immersed in an external spacetime whose local radius of curvature
is ${\cal R}$; this is assumed to be much larger than $M$, the mass of
the black hole. The tidal distortion is created by the Weyl curvature 
of the external spacetime, and the Schwarzschild metric acquires a
correction that can be expanded in powers of ${\cal R}^{-1}$. We
calculate all correction terms of order ${\cal R}^{-2}$ and 
${\cal R}^{-3}$; while the former were known previously
\cite{detweiler, poisson:04a} and used in Alvi's construction
\cite{alvi}, the latter are new. The metric is presented in a
light-cone coordinate system that penetrates the event horizon and  
possesses a clear geometrical meaning. These coordinates, which extend
over a neighborhood of the black hole that is large relative to $M$
but small relative to ${\cal R}$, allow an easy identification of the
perturbed event horizon. As an application we calculate the hole's
surface gravity and the rates at which the tidal interaction transfers
mass and angular momentum to the black hole. Because of space
limitations we present our results with very few derivations; a
forthcoming paper will supply the missing details.   

{\it Advanced coordinates based at a world line.} We begin with the
construction of a light-cone coordinate system in a neighborhood of a  
geodesic world line; the only assumption placed on the spacetime is
that its Ricci tensor vanishes in the neighborhood. This spacetime
will later play the role of the external spacetime, and the world line
will later be replaced by the world tube traced by the black hole. The 
quasi-Cartesian version of the advanced coordinates is denoted
$(v,x^a)$, and the quasi-spherical version $(v,r,\theta^A)$. The null
coordinate $v$ labels past light cones that are centered on the world
line, $r$ is an areal radius and an affine parameter on each cone's
null generators (with $r=0$ on the world line), and the angles
$\theta^A = (\theta,\phi)$ ($A = 2, 3$) are constant on each
generator. The coordinates $x^a$ ($a = 1, 2, 3$) are defined by 
$x^a := r \Omega^a(\theta^A)$ with $\Omega^a 
= (\sin\theta\cos\phi, \sin\theta\sin\phi, \cos\theta)$.  

Let $\gamma$ denote the geodesic world line, which is parameterized by
proper time $\tau$; its normalized tangent vector is $u^\mu$. Let 
$e^\mu_a(\tau)$ be a triad of orthonormal vectors (all orthogonal to 
$u^\mu$) which are parallel transported on the world line. Let
$\sigma(x,x')$ be half the squared geodesic distance between $x$ and
$x'$, and let $\sigma_{\alpha'}(x,x')$ denote its gradient with
respect to $x'$. The advanced coordinates are defined as follows. Let
$x'$ be the point on the world line at which $\tau$ takes the value
$v$. The set of events $x$ such that $\sigma(x,x') = 0$ and $x$ is in
the past of $x'$ defines the past light cone of $x'$. We assign the  
advanced-time coordinate $v$ to all these events. The spatial
coordinates of a point $x$ on the cone are then
$x^a = - \delta^{ab} e_b^{\alpha'}(v)\sigma_{\alpha'}(x,x')$. We 
introduce $r := (\delta_{ab} x^a x^b)^{1/2}$, which can be shown to be
an affine parameter on the null generators of the cone
$v=\mbox{constant}$. We also introduce $\Omega^a := x^a/r$, which may   
be parameterized by the two angles $\theta^A$. It can further be shown 
that $\theta^A$ are constant on the null generators.   

Tensors on $\gamma$ can be decomposed in the tetrad
$(u^\mu,e^\mu_a)$. For example, $C_{a0b0} :=
C_{\mu\lambda\nu\rho} e^\mu_a u^\lambda e^\nu_b u^\rho$ and
$C_{abc0} := C_{\mu\lambda\nu\rho} e^\mu_a e^\lambda_b e^\nu_c
u^\rho$ are frame components of the Weyl tensor evaluated on
$\gamma$. Similarly, $\dot{C}_{a0b0} := 
C_{\mu\lambda\nu\rho;\omega} e^\mu_a u^\lambda e^\nu_b u^\rho
u^\omega$ and $C_{a0b0|c} := C_{\mu\lambda\nu\rho;\omega} e^\mu_a
u^\lambda e^\nu_b u^\rho e^\omega_c$ are frame components of the
covariantly differentiated Weyl tensor. The 10 independent components 
of the Weyl tensor are conveniently related  to two
symmetric-tracefree (STF) frame tensors \cite{thorne-hartle},   
${\cal E}_{ab} := C_{a0b0}$ and ${\cal B}_{ab} := \frac{1}{2}
\varepsilon_{a}^{\ pq} C_{pqb0}$, where $\varepsilon_{abc}$ is the
permutation symbol (frame indices are freely manipulated with 
$\delta_{ab}$). The components of the differentiated Weyl tensor can
also be expressed in terms of STF frame tensors; these include 
$\dot{\cal E}_{ab}$ and $\dot{\cal B}_{ab}$ as well as 
${\cal E}_{abc} := \frac{1}{3}(C_{a0b0|c} + C_{c0a0|b} + C_{b0c0|a})$
and ${\cal B}_{abc} := \frac{1}{8} (\varepsilon_a^{\ pq} C_{pqb0|c} +  
\varepsilon_c^{\ pq} C_{pqa0|b} + \varepsilon_b^{\ pq}
C_{pqc0|a})$.  

These frame tensors allow us to define length scales
\cite{thorne-hartle} that characterize the tidal gravitational field
acting at $\gamma$. Let ${\cal R}$ be the local radius of curvature,
defined such that ${\cal E}_{ab} \sim {\cal B}_{ab} \sim 
{\cal R}^{-2}$. Let ${\cal L}$ be a scale of spatial inhomogeneity,
defined such that ${\cal E}_{abc} \sim {\cal B}_{abc} \sim 
{\cal R}^{-2} {\cal L}^{-1}$. And let ${\cal T}$ be a time scale
associated with changes in the tidal field, defined such that 
$\dot{\cal E}_{ab} \sim \dot{\cal B}_{ab} \sim {\cal R}^{-2}
{\cal T}^{-1}$. In the sequel we assume, for simplicity, that
${\cal R} \sim {\cal L} \sim {\cal T}$ and use ${\cal R}$ as the
fundamental length scale associated with the tidal field.      

The procedure to obtain the spacetime's metric in the coordinates  
$(v,x^a)$ or $(v,r,\theta^A)$ is detailed in
Ref.~\cite{poisson:04a}. The metric is expressed as an expansion in
powers of $r/{\cal R}$, which is assumed to be small; the coordinate
domain is therefore limited to a neighborhood of $\gamma$. The
appearance of the metric is simplified by forming irreducible
combinations involving the tidal fields ${\cal E}_{ab}(v)$, 
${\cal E}_{abc}(v)$, ${\cal B}_{ab}(v)$, ${\cal B}_{abc}(v)$, the unit
frame vector $\Omega^a = x^a/r$, the permutation symbol
$\varepsilon_{abc}$, and the frame metric $\delta_{ab}$. These
combinations are displayed in Table I.   

\begin{table}
\caption{Irreducible tidal fields. Each field is identified with a 
sans-serif superscript that specifies its multipole content. A field  
labeled with a ``{\sf q}'' is a quadrupole field, and one labeled
with an ``{\sf o}'' is an octupole field. The vector and tensor fields 
are all orthogonal to $\Omega^a$, and all tensor fields are symmetric
and tracefree.}  
\begin{ruledtabular}
\begin{tabular}{l}
$\EE = {\cal E}_{cd} \Omega^c \Omega^d$ \\ 
$\EE_a = (\delta_a^{\ c} - \Omega_a \Omega^c) 
{\cal E}_{cd} \Omega^d$ \\
$\EE_{ab} = 2(\delta_a^{\ c} - \Omega_a \Omega^c)(\delta_b^{\ d} 
-\Omega_b \Omega^d) {\cal E}_{cd} 
+ (\delta_{ab} - \Omega_a \Omega_b) \EE$ \\ 
$\BB_a = \varepsilon_{apq} \Omega^p {\cal B}^q_{\ c} \Omega^c$ \\ 
$\BB_{ab} = \varepsilon_{apq} \Omega^p {\cal B}^q_{\ c} 
(\delta^c_{\ b} - \Omega^c \Omega_b) + (a \leftrightarrow b)$ \\ 
$\EEE = {\cal E}_{cde} \Omega^c \Omega^d \Omega^e$ \\ 
$\EEE_a = (\delta_a^{\ c} - \Omega_a \Omega^c) 
{\cal E}_{cde} \Omega^d \Omega^e$ \\     
$\EEE_{ab} = 2(\delta_a^{\ c} - \Omega_a \Omega^c)(\delta_b^{\ d} 
-\Omega_b \Omega^d) {\cal E}_{cde} \Omega^e 
+ (\delta_{ab} - \Omega_a \Omega_b) \EEE$ \\ 
$\BBB_a = \frac{4}{3} \varepsilon_{apq} \Omega^p {\cal B}^q_{\ cd} 
\Omega^c \Omega^d$ \\
$\BBB_{ab} = \frac{4}{3} \varepsilon_{apq} \Omega^p {\cal B}^q_{\ cd}  
(\delta^c_{\ b} - \Omega^c \Omega_b) \Omega^d 
+ (a \leftrightarrow b)$
\end{tabular}
\end{ruledtabular}
\end{table}   

The quasi-Cartesian form of the metric is 
\begin{eqnarray} 
g_{vv} &=& -1 - r^2 \EE + \frac{1}{3} r^3 \EEd - \frac{1}{3} r^3 \EEE
+ O(4), 
\label{1} \\ 
g_{va} &=& \Omega_a - \frac{2}{3} r^2 \bigl( \EE_a - \BB_a \bigr) 
+ \frac{1}{3} r^3 \bigl( \EEd_a - \BBd_a \bigr) 
\nonumber \\ & & \mbox{}
- \frac{1}{4} r^3 \bigl( \EEE_a - \BBB_a \bigr) + O(4), 
\label{2} \\ 
g_{ab} &=& \delta_{ab} - \Omega_a \Omega_b 
- \frac{1}{3} r^2 \bigl( \EE_{ab} - \BB_{ab} \bigr) 
+ \frac{5}{18} r^3 \bigl( \EEd_{ab} - \BBd_{ab} \bigr) 
\nonumber \\ & & \mbox{}
- \frac{1}{6} r^3 \bigl( \EEE_{ab} - \BBB_{ab} \bigr) + O(4), 
\label{3}
\end{eqnarray} 
where $O(4)$ stands for ``a remainder that scales as 
${\cal R}^{-4}$.'' The quasi-spherical form of the metric is obtained
by performing the coordinate transformation $x^a = r
\Omega^a(\theta^A)$. Its nonvanishing components are $g_{vv}$ given by
Eq.~(\ref{1}), $g_{vr} = 1$, and   
\begin{eqnarray} 
g_{vA} &=& -\frac{2}{3} r^3 \bigl( \EE_A - \BB_A \bigr) 
+ \frac{1}{3} r^4 \bigl( \EEd_A - \BBd_A \bigr) 
\nonumber \\ & & \mbox{} 
- \frac{1}{4} r^4 \bigr( \EEE_A - \BBB_A \bigr) + O(4), 
\label{4} \\ 
g_{AB} &=& r^2 \Omega_{AB} 
- \frac{1}{3} r^4 \bigl( \EE_{AB} - \BB_{AB} \bigr) 
+ \frac{5}{18} r^5 \bigl( \EEd_{AB} - \BBd_{AB} \bigr) 
\nonumber \\ & & \mbox{} 
- \frac{1}{6} r^5 \bigr( \EEE_{AB} - \BBB_{AB} \bigr) + O(4).  
\label{5}
\end{eqnarray} 
Here $\Omega_{AB} := \delta_{ab} \Omega^a_A \Omega^b_B =
\mbox{diag}(1,\sin^2\theta)$, $\EE_A := \EE_a \Omega^a_A$,
$\EE_{AB} := \EE_{ab} \Omega^a_A \Omega^b_B$, and so on, with
$\Omega^a_A := \partial \Omega^a/\partial \theta^A$.  
The irreducible tidal fields can all be expanded in spherical
harmonics, and this decomposition is presented in Table II. According
to Eqs.~(\ref{1}), (\ref{4}), and (\ref{5}), $\sqrt{-g} 
= r^2\sin\theta[1 + O(4)]$, which implies that surfaces of constant
$v$ and $r$ have an area of $4\pi r^2[1 + O(4)]$.  

\begin{table*}
\caption{Spherical-harmonic decomposition of the irreducible tidal
fields. The first column lists $Y^{l{\sf m}}(\theta^A)$, the
real-valued spherical-harmonic functions that are used in the
decomposition. The abstract index $\sf m$ describes the dependence of  
these functions on $\phi$; for example $Y^{3,2s}$ is proportional to
$\sin 2\phi$. The second and third columns define 
$\EE_{\sf m}(v)$, $\EEE_{\sf m}(v)$, $\BB_{\sf m}(v)$, 
$\BBB_{\sf m}(v)$, the harmonic components of the tidal fields, in
terms of their frame components ${\cal E}_{ab}(v)$, 
${\cal E}_{abc}(v)$, ${\cal B}_{ab}(v)$, ${\cal B}_{abc}(v)$. The
fourth column displays the spherical-harmonic decompositions of the
quadrupole and octupole tidal fields. The vectorial spherical
harmonics are $Y^{l {\sf m}}_{A} := D_A Y^{l {\sf m}}$ and 
$X^{l {\sf m}}_{A} := -\varepsilon_A^{\ B} D_B Y^{l {\sf m}}$, where
$D_A$ is the covariant derivative operator on the unit two-sphere 
($D_A \Omega_{BC} = 0$) and $\varepsilon_{AB}$ is the Levi-Civita
tensor ($\varepsilon_{\theta\phi} = \sin\theta$; indices raised
with $\Omega^{AB}$, the inverse two-sphere metric). The tensorial
spherical harmonics are $Y^{l {\sf m}}_{AB} := [D_A D_B 
+ \frac{1}{2} l(l+1) \Omega_{AB}] Y^{l {\sf m}}$ and 
$X^{l {\sf m}}_{AB} = -\frac{1}{2}(\varepsilon_{A}^{\ C} D_{B} 
+ \varepsilon_{B}^{\ C} D_{A}) D_C Y^{l {\sf m}}$.}
\begin{ruledtabular} 
\begin{tabular}{llll}
$Y^{2,0} = -(3\cos^2\theta-1)$ & 
$\EE_0 = \frac{1}{2}({\cal E}_{11} + {\cal E}_{22})$ & 
$\BB_0 = \frac{1}{2}({\cal B}_{11} + {\cal B}_{22})$ & 
$\EE = \sum_{\sf m} \EE_{\sf m} Y^{2 {\sf m}}$ \\ 
$Y^{2,1c} = 2\sin\theta\cos\theta\cos\phi$ & 
$\EE_{1c} = {\cal E}_{13}$ &
$\BB_{1c} = {\cal B}_{13}$ &
$\EE_A = \frac{1}{2} \sum_{\sf m} \EE_{\sf m} Y^{2 {\sf m}}_{A}$ \\  
$Y^{2,1s} = 2\sin\theta\cos\theta\sin\phi$ & 
$\EE_{1s} = {\cal E}_{23}$ &
$\BB_{1s} = {\cal B}_{23}$ & 
$\EE_{AB} = \sum_{\sf m} \EE_{\sf m} Y^{2 {\sf m}}_{AB}$ \\ 
$Y^{2,2c} = \sin^2\theta \cos 2\phi$ & 
$\EE_{2c} = \frac{1}{2} ({\cal E}_{11} - {\cal E}_{22})$ & 
$\BB_{2c} = \frac{1}{2} ({\cal B}_{11} - {\cal B}_{22})$ & 
$\BB_A = \frac{1}{2} \sum_{\sf m} \BB_{\sf m} X^{2 {\sf m}}_{A}$ \\   
$Y^{2,2s} = \sin^2\theta \sin 2\phi$ & 
$\EE_{2s} = {\cal E}_{12}$ &
$\BB_{2s} = {\cal B}_{12}$ &
$\BB_{AB} = \sum_{\sf m} \BB_{\sf m} X^{2 {\sf m}}_{AB}$ \\  
$Y^{3,0} = -(5\cos^3\theta-3\cos\theta)$ & 
$\EEE_0 = \frac{1}{2}({\cal E}_{113} + {\cal E}_{223})$ &   
$\BBB_0 = \frac{2}{3}({\cal B}_{113} + {\cal B}_{223})$ &
\\ 
$Y^{3,1c} = -\frac{3}{2} \sin\theta(5\cos^2\theta-1)\cos\phi$ & 
$\EEE_{1c} = \frac{1}{2}({\cal E}_{111} + {\cal E}_{122})$ & 
$\BBB_{1c} = \frac{2}{3}({\cal B}_{111} + {\cal B}_{122})$ &
\\
$Y^{3,1s} = -\frac{3}{2} \sin\theta(5\cos^2\theta-1)\sin\phi$ & 
$\EEE_{1s} = \frac{1}{2}({\cal E}_{112} + {\cal E}_{222})$ & 
$\BBB_{1s} = \frac{2}{3}({\cal B}_{112} + {\cal B}_{222})$ & 
$\EEE = \sum_{\sf m} \EEE_{\sf m} Y^{3 {\sf m}}$ \\   
$Y^{3,2c} = 3\sin^2\theta\cos\theta\cos 2\phi$ &
$\EEE_{2c} = \frac{1}{2}({\cal E}_{113} - {\cal E}_{223})$ & 
$\BBB_{2c} = \frac{2}{3}({\cal B}_{113} - {\cal B}_{223})$ &
$\EEE_A = \frac{1}{3} \sum_{\sf m} \EEE_{\sf m} Y^{3 {\sf m}}_{A}$ \\   
$Y^{3,2s} = 3\sin^2\theta\cos\theta\sin 2\phi$ &
$\EEE_{2s} = {\cal E}_{123}$ & 
$\BBB_{2s} = \frac{4}{3} {\cal B}_{123}$ &
$\EEE_{AB} = \frac{1}{3} \sum_{\sf m} \EEE_{\sf m} Y^{3 {\sf m}}_{AB}$
\\  
$Y^{3,3c} = \sin^3\theta\cos 3\phi$ & 
$\EEE_{3c} = \frac{1}{4}({\cal E}_{111} - 3{\cal E}_{122})$ &
$\BBB_{3c} = \frac{1}{3}({\cal B}_{111} - 3{\cal B}_{122})$ &
$\BBB_A = \frac{1}{3} \sum_{\sf m} \BBB_{\sf m} X^{3 {\sf m}}_{A}$ \\  
$Y^{3,3s} = \sin^3\theta\sin 3\phi$ & 
$\EEE_{3s} = \frac{1}{4}(3{\cal E}_{112} - {\cal E}_{222})$ &
$\BBB_{3s} = \frac{1}{3}(3{\cal B}_{112} - {\cal B}_{222})$ &
$\BBB_{AB} = \frac{1}{3} \sum_{\sf m} \BBB_{\sf m} X^{3 {\sf m}}_{AB}$
\end{tabular}
\end{ruledtabular}
\end{table*}   

{\it Metric of a tidally distorted black hole.} The metric of an
isolated, nonrotating black hole of mass $M$ can be put in the form  
\begin{equation}
g^0_{\alpha\beta} dx^\alpha dx^\beta 
= -f\, dv^2 + 2\, dvdr + r^2 \Omega_{AB}\, d\theta^A d\theta^B, 
\label{6}
\end{equation}
where $f := 1-2M/r$ and $\Omega_{AB}\, d\theta^A d\theta^B
= d\theta^2 + \sin^2\theta\, d\phi^2$. The coordinates
$(v,r,\theta^A)$ have the same meaning as in the preceding section:
$v$ labels past light cones, $r$ is an areal radius and an affine
parameter on each cone's null generators, and $\theta^A$ are constant
on each generator. The metric of Eq.~(\ref{6}) approaches the
Minkowski metric in the asymptotic regime $r \gg 2M$.  

A tidally distorted black hole is described by a metric that is a
perturbed version of Eq.~(\ref{6}). In the asymptotic regime $r \gg
2M$ (keeping $r \ll {\cal R}$, which implies that $M$ must be much
smaller than ${\cal R}$) this metric must approach the metric of
Eqs.~(\ref{1}), (\ref{4}), and (\ref{5}) instead of the Minkowski
metric. This observation motivates the following ansatz for the
perturbed metric:   
\begin{eqnarray} 
g_{vv} &=& -f - r^2 e_1 \EE + \frac{1}{3} r^3 e_2 \EEd 
- \frac{1}{3} r^3 e_3 \EEE + O(4), \quad 
\label{7} \\ 
g_{vr} &=& 1, 
\label{8} \\
g_{vA} &=& -\frac{2}{3} r^3 \bigl( e_4 \EE_A - b_4 \BB_A \bigr) 
+ \frac{1}{3} r^4 \bigl( e_5 \EEd_A - b_5 \BBd_A \bigr) 
\nonumber \\ & & \mbox{} 
- \frac{1}{4} r^4 \bigr( e_6 \EEE_A - b_6 \BBB_A \bigr) + O(4), 
\label{9} \\ 
g_{AB} &=& r^2 \Omega_{AB} 
- \frac{1}{3} r^4 \bigl( e_7 \EE_{AB} - b_7 \BB_{AB} \bigr) 
\nonumber \\ & & \mbox{} 
+ \frac{5}{18} r^5 \bigl( e_8 \EEd_{AB} - b_8 \BBd_{AB} \bigr) 
\nonumber \\ & & \mbox{} 
- \frac{1}{6} r^5 \bigr( e_9 \EEE_{AB} - b_9 \BBB_{AB} \bigr) + O(4).   
\label{10}
\end{eqnarray} 
The radial functions $e_n(r)$, $b_n(r)$ must be determined by
integrating the Einstein field equations; they are independent of
${\cal R}$, they are required to be well-behaved at $r=2M$, and they
must approach unity when $r \gg 2M$. The perturbed metric of
Eqs.~(\ref{7})--(\ref{10}) is expressed in a gauge that preserves the 
geometrical meaning of the coordinates $(v,r,\theta^A)$; this gauge is
both physically meaningful and technically simple. The gauge freedom
is not exhausted by adopting the ansatz of Eqs.~(\ref{7})--(\ref{10});
refinements can be made when solving the field equations.       

The metric of Eqs.~(\ref{7})--(\ref{10}) is used to compute
the Einstein tensor, which is then expanded in powers of 
${\cal R}^{-1}$. There is no contribution at orders ${\cal R}^0$ and 
${\cal R}^{-1}$, but we examine carefully the contributions at orders
${\cal R}^{-2}$ and ${\cal R}^{-3}$. At order ${\cal R}^{-2}$ the
Einstein tensor decouples into spherical-harmonic components that can
be handled separately, and it decouples further into even-parity terms  
proportional to $\EE_{\sf m}$ and odd-parity terms proportional to
$\BB_{\sf m}$. The even-parity sector gives rise to a coupled set of
ordinary differential equations for $\{e_1, e_4, e_7\}$, while the 
odd-parity sector determines $\{b_4, b_7\}$. General solutions are
easily found, and they are constrained by demanding that the radial
functions be well-behaved at $r=2M$ and approach unity as 
$r \to \infty$. This leaves a number of integration constants
undetermined, and those can be freely assigned; this represents a
refinement of the gauge, and we adjust the constants so that a maximum
number of radial functions vanish at $r=2M$.  

This procedure can be completed at order ${\cal R}^{-3}$. Here the
octupole terms (proportional to $\EEE_{\sf m}$ and $\BBB_{\sf m}$)
decouple from the quadrupole terms (proportional to $\EEd_{\sf m}$ and
$\BBd_{\sf m}$), and as before the even-parity terms decouple from the
odd-parity terms. We thus obtain coupled sets of equations for 
$\{e_2,e_5,e_8\}$, $\{b_5,b_8\}$, $\{e_3,e_6,e_9\}$, and 
$\{b_6,b_9\}$. These are integrated using the same techniques as for
the ${\cal R}^{-2}$ part of the metric. The radial functions are
listed in Table III.              

\begin{table}
\caption{Radial functions that appear in the metrics of
Eqs.~(\ref{7})--(\ref{10}), expressed in terms of $x
:= r/(2M)$ and $f := 1-2M/r$. At $r=2M$ we have $e_7 =
\frac{1}{2}$, $e_9 = \frac{1}{10}$, $b_7 = -\frac{1}{2}$, 
and $b_9 = -\frac{1}{10}$, with all other functions vanishing.}    
\begin{ruledtabular}
\begin{tabular}{l}
$e_1 = f^2$ \\ 
$e_2 = f[ 1 + \frac{1}{4x} (5 + 12\ln x) - \frac{3}{4x^2} (9 + 4\ln x)
+ \frac{7}{4x^3} + \frac{3}{4x^4}]$ \\ 
$e_3 = f^2(1 - \frac{1}{2x})$ \\ 
$e_4 = f$ \\
$e_5 = f[ 1 + \frac{1}{6x}(13 + 12\ln x) - \frac{5}{2x^2} 
- \frac{3}{2x^3} - \frac{1}{2x^4}]$ \\ 
$e_6 = f(1 - \frac{2}{3x})$ \\ 
$e_7 = 1 - \frac{1}{2x^2}$ \\ 
$e_8 = 1 + \frac{2}{5x}(4 + 3\ln x) - \frac{9}{5x^2} 
- \frac{1}{5x^3}(7 + 3\ln x) + \frac{3}{5x^4}$ \\ 
$e_9 = f + \frac{1}{10x^3}$ \\ 
$b_4 = f$ \\
$b_5 = f[ 1 + \frac{1}{6x}(7 + 12\ln x) - \frac{3}{2x^2} 
- \frac{1}{2x^3} - \frac{1}{6x^4}]$ \\ 
$b_6 = f(1 - \frac{2}{3x})$ \\ 
$b_7 = 1 - \frac{3}{2x^2}$ \\ 
$b_8 = 1 + \frac{1}{5x}(5 + 6\ln x) - \frac{9}{5x^2} 
- \frac{1}{5x^3}(2 + 3\ln x) + \frac{1}{5x^4}$ \\  
$b_9 = f - \frac{1}{10x^3}$ 
\end{tabular}
\end{ruledtabular}
\end{table}   

The quasi-Cartesian form of the perturbed metric is obtained by
inserting $e_n$ and $b_n$ at appropriate places in
Eqs.~(\ref{1})--(\ref{3}); for example, the coefficient of  
$\EE_a$ in $g_{va}$ becomes $-\frac{2}{3} r^2 e_4$. The complete 
determination of the metric requires the specification of the tidal
fields ${\cal E}_{ab}(v)$, ${\cal E}_{abc}(v)$, ${\cal B}_{ab}(v)$,
and ${\cal B}_{abc}(v)$. These characterize the black hole's local 
environment, and they are determined by matching the hole's local
metric to a metric that describes a larger portion of the
spacetime \cite{thorne-hartle,death}. For example, the global metric
could be the post-Newtonian metric of a two-body system, and Alvi
\cite{alvi} has shown how to carry out the matching procedure in this
context.        

{\it The perturbed horizon.} To locate the event horizon of the
tidally distorted black hole we note first that the vector $r_\alpha
:= \partial_\alpha r$ is normal to any surface of constant $r$, and
that it is null when $r = 2M$: $g^{\alpha\beta} r_\alpha r_\beta =
O(4)$. The generators of this null surface have $r^\alpha$ as their
tangent vector, and a simple calculation reveals that $r^\alpha 
= t^\alpha + O(4)$, where $t^\alpha (\partial/\partial x^\alpha) 
:= \partial/\partial v$. The generators are therefore parameterized by
$v$ and they move with constant values of $\theta^A$ (up to
corrections of order ${\cal R}^{-4}$). That $r=2M$ is in fact the
event horizon is confirmed by a theorem established in
Ref.~\cite{poisson:04b}: {\it Let $g_{\alpha\beta}$ be the metric of a 
perturbed Schwarzschild black hole, let $h_{\alpha\beta} := 
g_{\alpha\beta} - g^0_{\alpha\beta}$ be the metric perturbation, and
let $t^\alpha$ be tangent to the null generators of the perturbed
horizon; in a coordinate system in which $h_{\alpha\beta} t^\beta = 0$
on the horizon, the horizon has the same coordinate description in the
perturbed and unperturbed spacetimes.} In other words, the gauge
conditions $h_{\alpha v} = 0$ at $r=2M$ guarantee that the event
horizon is located at $r=2M$ in both spacetimes. Because the metric of 
Eqs.~(\ref{7})--(\ref{10}) satisfies these conditions to order 
${\cal R}^{-4}$, we conclude that the surface $r=2M[1+O(4)]$ is the
event horizon of the tidally distorted black hole.   

The surface gravity $\kappa$ of the perturbed horizon is defined
by the statement $t^\alpha_{\ ;\beta} t^\beta = \kappa t^\alpha$,
which holds on $r=2M$ and expresses the fact that the integral curves
of $t^\alpha$ are null geodesics. A straightforward calculation gives  
\begin{equation} 
\kappa = \frac{1}{4M}\biggl[ 1 + \frac{16}{3} M^3 \dot{\cal E}_{ab}(v)
\Omega^a \Omega^b + O(4) \biggr]. 
\label{13}
\end{equation} 
The expansion of the congruence of null generators is calculated as 
$\Theta = (g^{\alpha\beta} + t^\alpha l^\beta + l^\alpha t^\beta)
t_{\alpha;\beta}$, where $l_\alpha := -\partial_\alpha v$ is the other
null normal to the event horizon. A straightforward calculation
reveals that $\Theta = O(M^4/{\cal R}^5)$. This result implies that
within the accuracy of the calculation, each cross section
$v=\mbox{constant}$ of the event horizon is in fact an apparent
horizon.  

The slow growth of the event horizon is not revealed by a direct
examination of the horizon's geometry. It can, however, be calculated 
using the techniques introduced in Ref.~\cite{poisson:04b}. Such a
calculation reveals that the averaged rate at which the black hole
increases its mass because of tidal heating is given by 
\begin{eqnarray} 
\biggl\langle \frac{dM}{dv} \biggr\rangle &=& 
\frac{16}{45} M^6 \bigl\langle \dot{\cal E}_{ab} \dot{\cal E}^{ab} 
+ \dot{\cal B}_{ab} \dot{\cal B}^{ab} \bigr\rangle 
\nonumber \\ & & \mbox{} 
+ \frac{16}{4725} M^8 \biggl\langle \dot{\cal E}_{abc} 
\dot{\cal E}^{abc} + \frac{16}{9} \dot{\cal B}_{abc} 
\dot{\cal B}^{abc} \biggr\rangle \qquad
\label{14}
\end{eqnarray} 
with a remainder of order $(M/{\cal R})^9$. The rate at which the
horizon's surface area grows is then $\langle dA/dv \rangle = 
(8\pi/\kappa) \langle dM/dv \rangle$. The tidal interaction also
transfers angular momentum to the black hole; the averaged rate of
change of its angular-momentum vector is given by   
\begin{eqnarray} 
\biggl\langle \frac{d J^a}{dv} \biggr\rangle &=& -\frac{32}{45} M^6 
\varepsilon^{a}_{\ bc} \bigl\langle \dot{\cal E}^b_{\ d} {\cal E}^{cd}
+ \dot{\cal B}^b_{\ d} {\cal B}^{cd} \bigr\rangle 
\nonumber \\ & & \hspace*{-20pt} \mbox{} 
- \frac{16}{1575} M^8 \varepsilon^{a}_{\ bc} \biggl\langle 
\dot{\cal E}^b_{\ de} {\cal E}^{cde} + \frac{16}{9} 
\dot{\cal B}^b_{\ de} {\cal B}^{cde} \biggr\rangle \qquad 
\label{15}
\end{eqnarray} 
with a remainder of order $M^9/{\cal R}^8$.  

{\it Conclusion.} The metric of an arbitrary vacuum spacetime was
expressed in a light-cone coordinate system centered on a geodesic
world line; the metric was expanded to third order in powers of 
$r/{\cal R}$, which is assumed to be small in the coordinate
domain. This provided asymptotic conditions on the metric of a tidally
distorted, nonrotating black hole, which was obtained by integrating
the equations of black-hole perturbation theory to order 
${\cal R}^{-3}$. The coordinates allow an easy identification of the
event horizon, and the effects of the tidal interaction on its
physical properties were explored. This work was supported
by the Natural Sciences and Engineering Research Council of Canada.

\end{document}